\documentclass[aps,pre,twocolumn,nofootinbib]{revtex4-2}
\usepackage{graphicx}
\usepackage{dcolumn}
\usepackage{bm,bbm}
\usepackage{amsthm,amsmath,amssymb}
\usepackage{algorithm}
\usepackage{algpseudocode}
\usepackage[english]{babel}

\usepackage{xcolor}
\newcommand{\Dt}{{\triangle t}}

\begin{document}

\title{Bounded-confidence opinion models with random-time interactions}

\author{Weiqi Chu}
\affiliation{University of Massachusetts Amherst, Amherst, Massachusetts, USA}
\author{Mason A. Porter}
\affiliation{University of California, Los Angeles, Los Angeles, California, USA}
\affiliation{Santa Fe Institute, Santa Fe, New Mexico, USA}

\date{\today}


\begin{abstract}

    In models of opinion dynamics, agents interact with each other and can change their opinions as a result of those interactions. 
    One type of opinion model is a bounded-confidence model (BCM), in which opinions take continuous values and interacting agents compromise their opinions with each other if their opinions are sufficiently similar.
    In studies of BCMs, researchers typically assume that interactions between agents occur at deterministic times. 
    This assumption neglects an inherent element of randomness in social interactions, and it is desirable to account for it.
    In this paper, we study BCMs on networks and allow agents to interact at random times. 
    To incorporate random-time interactions, we use renewal processes to determine social-interaction event times, which can follow arbitrary interevent-time distributions (ITDs).
    We establish connections between these random-time-interaction BCMs and deterministic-time-interaction BCMs. 
    We analyze the quantitative impact of ITDs on the transient dynamics of BCMs and derive approximate master equations for the time-dependent expectations of the BCM dynamics.
    We find that BCMs with Markovian ITDs have consistent statistical properties (in particular, they have the same expected time-dependent opinions) when the ITDs have the same mean but that the statistical properties of BCMs with non-Markovian ITDs depend on the type of ITD even when the ITDs have the same mean. 
    Additionally, we numerically examine the transient and steady-state dynamics of our models with various ITDs on different networks and compare their expected order-parameter values and expected convergence times.
\end{abstract}

\maketitle


\section{Introduction} \label{zero}

On social-media platforms, individuals engage in regular and frequent exchanges of opinions, and people's views and how those views change play a pivotal role in shaping societal discourse~\cite{bak2021}. 
The study of opinion dynamics --- which involves the intersection of the social and behavioral sciences, mathematics, complex systems, and other areas --- has emerged as a vibrant research area that aims to determine the mechanisms that govern the formation, evolution, and dissemination of opinions in human (and animal) societies~\cite{starnini2025,caldarelli2025,noorazar2020,olsson2023,castellano2009statistical,sen2014sociophysics,jusup2022}. At its core, the study of opinion dynamics concerns how beliefs, attitudes, and perceptions evolve with time through agreement, compromise, persuasion, imitation, and conflict.
Studying such dynamics is crucial to understanding both (1) the emergence of consensus, polarization, and fragmentation and (2) the resilience of diverse opinions in societies, especially in the modern ecosystem of increasingly interconnected and digital communication environments~\cite{amelkin2017polar,mathias2017energy,grabowski2009opinion}.

Researchers have studied many types of opinion models~\cite{noorazar2020,starnini2025,caldarelli2025}.
In opinion models, agents adjust their opinions based on their interactions with other agents. 
Their opinions can update either in discrete time or in continuous time.
In opinion models with discrete-time updates, time progresses through a sequence of discrete steps. Examples of discrete-time opinion models include voter models \cite{sood2005voter}, DeGroot consensus models \cite{degroot1974reaching}, and bounded-confidence models (BCMs) \cite{deffuant2000mixing,hegselmann2002opinion}.
Opinion models with discrete-time updates are straightforward to implement for numerical simulations, and one can readily incorporate various features (such as parameter adaptivity \cite{li2023bounded}) into such models.
In opinion models with continuous-time interactions and hence continuous-time updates, agents continuously adjust their opinions at rates that are influenced by factors such as whether they have friendly or hostile relationships with their neighbors~\cite{altafini2012dynamics} and the difference between their opinions and the opinions of their neighbors~\cite{blondel2010continuous,brooks2024}. Another prominent type of model with continuous-time interactions is density-based opinion models \cite{ben2003bifurcations}, which consider the collective evolution of opinions in a large population and often are described by integro-differential equations.

Several researchers have highlighted the importance of incorporating randomness into opinion models to accurately capture the probabilistic nature of human interactions~\cite{diekmann1993cooperation,ajzen2020theory}. One can incorporate randomness in the structure of social and communication ties between agents by using random networks, such as configuration models, stochastic block models (SBMs), and their generalizations \cite{newman2018networks}. 
Additionally, one can use tie-decay networks \cite{sugishita2021opinion} (which distinguish between communication processes and underlying social ties) and activity-driven networks \cite{perra2012} (which also incorporate randomness in the interactions between agents) to incorporate randomness in communication. Another way to introduce randomness in a model is to incorporate noise and employ a stochastic differential equation (SDE) to describe opinion evolution \cite{goddard2022noisy,pineda2009noisy}.
One can also incorporate probabilistic components into the decision-making process of agents during opinion updates~\cite{redner2019reality,deffuant2000mixing,fernandez2011update} either by choosing a random pair of agents to interact at each time step \cite{deffuant2000mixing} or by allowing agents to choose probabilistically between multiple opinion-update rules \cite{fernandez2011update}. 
See \cite{schawe2021network} for a quantitative study of how randomness in the structure of specific network models (including Erd\H{os}--R\'{e}nyi (ER) graphs and Barab\'asi--Albert (BA) graphs) influences steady-state features, phase transitions, consensus formation, and finite-size effects in the Hegselmann--Krause (HK) BCM.

Temporal stochasticity is another form of randomness that is relevant to opinion models, but it is often overlooked. Existing opinion models typically treat time as deterministic and neglect the temporal stochasticity that is inherent in social interactions. 
In the present paper, we model social interactions using renewal processes~\cite{feller-renewal}. A renewal process consists of a sequence of random events, and the time between consecutive events follows a desired interevent-time distribution (ITD). 
By employing renewal processes, we are able to study non-Markovian dynamical processes, which arise frequently in human dynamics, including in financial markets~\cite{scalas2006application}, the spread of infectious diseases~\cite{masuda2013predicting}, e-mail traffic~\cite{barabasi2005origin}, and opinion dynamics~\cite{chu2023nonmarkovian}.
For concreteness, we frame our discussion in the context of BCMs~\cite{lorenz2007continuous,bernardo2024}. We consider both HK models (in which agents update their opinions synchronously) and Deffuant--Weisbuch (DW) models (in which agents update their opinions asynchronously). BCMs have been studied extensively by physicists, mathematicians, and others. 
For results about consensus formation, convergence, and opinion clustering in BCMs, see \cite{blondel2009krause,dittmer2001consensus} for HK models and \cite{lorenz2005stabilization,chen2024convergence,meng2018opinion} for DW models.
We discuss two approaches to integrate temporal stochasticity into BCMs, and we investigate the effects of stochasticity on the convergence of opinions, the formation of opinion clusters, and the transient dynamics of opinions.
We establish connections between our opinion models and classical BCMs, and we approximate the expected dynamics of non-Markovian opinion dynamics using BCMs with interactions at deterministic times.


Our paper proceeds as follows. 
In Section \ref{one}, we discuss single-process BCMs, in which a single renewal process dictates all the interaction times of all agents. We explore these BCMs with both synchronous and asynchronous update rules by examining properties such as expected dynamics, convergence, and other aspects for different ITDs. 
In Section \ref{two}, we discuss multiple-process BCMs, where independent renewal processes govern the interaction times between each pair of agents. 
We derive the expected dynamics for Markovian BCMs in this framework, and we use a Gillespie algorithm to efficiently simulate event times for non-Markovian BCMs.
In Section \ref{three}, we conclude and discuss future directions.
Our code is available at \url{https://bitbucket.org/chuwq/bounded_confidence_models_with/src/main/}.


\section{Single-process BCMs} \label{one}


\subsection{Random-time interactions}

Consider an unweighted and directed network (i.e., graph) $G = (V, E)$, where $V = \{1,2,\ldots,N\}$ is the set of nodes (i.e., agents) and $E = \{e_{ij}\}$ is the set of edges (i.e., social ties between agents). The directed edge $e_{ij}$ starts at agent $j$ and ends at agent $i$. Each agent $i$ has a scalar continuous-valued opinion $x_i(t)$. When $e_{ij} = 1$, agent $j$ can potentially influence agent $i$'s opinion.
In a classical BCM~\cite{bernardo2024,deffuant2000mixing,hegselmann2002opinion}, time is deterministic and takes discrete values, with social interactions and opinion updates occurring in intervals of duration $\Delta t$. For convenience, researchers often set $\Delta t = 1$.

Let $R(t)$ be a renewal process, which is a stochastic process that models a sequence of events that occur randomly in time \cite{feller-renewal}. Let $\mathcal{T} = \left\{t_0,t_1,t_2,\ldots\right\}$ be the sequence of event times in the renewal process $R(t)$. We set $t_0 = 0$ as the starting time of the renewal process. The time increments (i.e., \textit{interevent times}) $t_{k + 1} - t_k$ constitute a sequence of independent and identically distributed (IID) random variables with finite expected values. Because $t_{k + 1} > t_k$ for all $k$, the time increments are positive. Let $\psi(t)$ denote the probability density function (PDF) of the IID random variables $t_{k + 1} - t_k$.
It is common to refer to this PDF as an \emph{interevent-time distribution}~\cite{hoffmann2012generalized,speidel2015}.
In this section, we suppose that a single renewal process determines the interaction times between the agents in a network.


\subsection{Synchronous and asynchronous opinion-update rules} \label{sec: HK}

The HK model \cite{hegselmann2002opinion,krause2000discrete} is a discrete-time BCM with a synchronous opinion-update rule. That is, all agents update their opinions simultaneously. Let\footnote{In \cite{hegselmann2002opinion,krause2000discrete}, the set of neighbors of each agent $i$ is $\mathcal{N}_i(t) = \{i\} \cup \left\{j: ~e_{ij}\in E \,\,\text{~and~}\,\, |x_i(t) - x_j(t)| \leq c\right\}$. To be consistent with the strict inequality in the classical DW BCM \cite{deffuant2000mixing}, we instead use a strict inequality.} 
\begin{equation} \label{eq: neighbors}
	\mathcal{N}_i(t) = \{i\} \cup \left\{j: ~e_{ij}\in E \,\,\text{~and~}\,\, |x_i(t) - x_j(t)| < c\right\}
\end{equation}
be the set of neighbors of agent $i$ (including $i$ itself) with which it interacts at time $t$. The parameter $c$ is the confidence bound.
In each time step, the opinion of each agent $i$ updates through the rule
\begin{equation} \label{classical}
    x_i(t + \Dt) = \frac{\sum_{j\in\mathcal{N}_i(t)} x_j(t)}{|\mathcal{N}_i(t)|}\,, \quad t = 0,\,\Dt,\,2\Dt,\,\ldots \, \,. 
\end{equation}
We extend the HK BCM to a continuous-time model with interactions on directed graphs at random times.
Agents update their opinions synchronously when an event occurs in the renewal process $R(t)$. The opinion-update rule is thus
\begin{equation}\label{eq: HK-random}
    x_i(t) = \frac{\sum_{j\in\mathcal{N}_i(t_-)} x_j(t_-)}{|\mathcal{N}_i(t_-)|}\,, \quad  t \in \mathcal{T} = \{t_1,t_2,\ldots\}\,,
\end{equation}
where $t_- = \lim_{\epsilon \rightarrow 0}[t - \epsilon]$ (with $\epsilon > 0$) denotes the time that is instantaneously before time $t$. Therefore, $x_j(t_-)$ is the opinion of agent $j$ right before it updates its opinion at time $t$.
Unless an event occurs at time $t \in \mathcal{T}$, the opinions of all agents stay the same.
We refer to the BCM with the opinion-update rule \eqref{eq: HK-random} as a \emph{single-process BCM} with synchronous updates. 
When the ITD $\psi$ is the Dirac delta distribution (i.e., $\psi(t) = \delta(t-\Dt)$), the update rule \eqref{eq: HK-random} reduces to the update rule \eqref{classical} in the classical HK BCM \cite{hegselmann2002opinion}.


Deffuant et al.~\cite{deffuant2000mixing} introduced a discrete-time BCM with an asynchronous opinion-update rule. At each discrete time step, one selects a pair of agents uniformly at random and updates their opinions to the mean of their opinions (or, more generally, to opinions that are closer to the mean) if their opinion difference is smaller than a confidence bound $c$.
This model, which is the DW BCM, was proposed in the context of undirected graphs. We extend the DW BCM to a directed DW BCM. 
In this directed DW model, at time step $t$, one selects an edge $e_{ij}$ uniformly at random and updates the opinion of agent $i$ with the rule\footnote{In the classical DW model \cite{deffuant2000mixing}, one chooses a random edge and potentially updates the opinions of its two attached nodes. By contrast, in the present paper, we choose a random edge and then potentially update the opinion of only one node. A third option, which was employed in \cite{li2023}, is to first choose a random node, then randomly choose one of its neighboring nodes to interact with it, and then potentially update the opinions of both nodes.}
\begin{equation}\label{eq: DW_classical}
	x_i(t + \Dt) = 
		\begin{cases}
			 \frac{1}{2} \left[x_j(t) + x_i(t)\right] &  \text{if} \,\, |x_i(t) - x_j(t)| < c  \\
			    x_i(t) & \text{otherwise}\,.
\end{cases}
\end{equation}
The opinions of all other agents stay the same. The time $t$ takes values from the set $\left\{0,\,\Dt,\,2\Dt,\,\ldots\right\}$. 
Unlike in the classical DW model \cite{deffuant2000mixing}, we update the opinion of only agent $i$ rather than updating the opinions of both agent $i$ and agent $j$. Unidirectional interactions and influence occur in many social and biological systems \cite{guttal2010social,hemelrijk1990models}. For example, individuals may update their opinions by reading other individuals' social-media posts without commenting or otherwise signaling their engagement with those posts.

We generalize the directed DW BCM \eqref{eq: DW_classical} to a continuous-time model by allowing interactions at random times. At time $t \in \mathcal{T}$, where $\mathcal{T}$ is the set of event times of a renewal process $R(t)$, we select an edge 
$e_{ij}$ uniformly at random and update the opinion of agent $i$ with the rule
\begin{equation}\label{eq: DW-random}
	x_i(t) =
		\begin{cases}
		    \frac{1}{2} \left[x_j(t_-) + x_i(t_-)\right]  & \text{if} \,\, |x_i(t_-) - x_j(t_-)| < c \\
		     x_i(t_-) &\text{otherwise}\,.
		\end{cases}
\end{equation}
The opinions of all other agents stay the same. Additionally, unless an event occurs at time $t \in \mathcal{T}$, the opinions of all agents stay the same. 
The BCM with the opinion-update rule \eqref{eq: DW-random} is a single-process BCM with asynchronous updates. 
When the ITD $\psi$ is the Dirac delta distribution, the update rule \eqref{eq: DW-random} reduces to the update rule \eqref{eq: DW_classical} in the directed DW BCM.

In the random-time BCMs with synchronous \eqref{eq: HK-random} and asynchronous \eqref{eq: DW-random} update rules, the agent opinions converge almost surely to isolated opinion clusters (i.e., maximal sets of agents with the same opinion value) that differ by at least the confidence bound $c$. This is a direct consequence of Lorenz's stability theorem \cite{lorenz2005stabilization}.


\subsection{Exact and approximate dynamics of the expected opinions}
\label{sec: SP-analysis}

Let ${\bm x}(t) = (x_1(t),\ldots,x_N(t)) \in \mathbb{R}^N$ be the time-dependent opinion vector of the single-process BCM \eqref{eq: HK-random} or \eqref{eq: DW-random}. 
The randomness in $\bm x(t)$ arises from the interaction times, the selection of edges in the asynchronous-update model \eqref{eq: DW-random}, and potentially random initial opinions. These three sources of randomness are independent of each other. In the rest of this section, we fix the initial opinion vector $\bm x_0$ and investigate how the other two sources of randomness influence the dynamics of the expected opinions.
We also examine how the choice of ITD influences the dynamics of the expected opinions in single-process BCMs with the synchronous update rule \eqref{eq: HK-random} and the asynchronous update rule \eqref{eq: DW-random}.

Let $u_k(t)$ denote the probability that the renewal process $R(t)$ has $k$ events in the time interval $[0,t]$. With the ITD $\psi$, we have $u_0(t) = 1 - \int_0^t \psi(\tau)\mathrm{d}\tau$. If $k + 1$ events occur in the time interval $[0,t]$, then $k$ events occur in the time interval $[0,t - \tau]$ and $1$ event occurs in the time interval $(t - \tau,t]$ for some $\tau\in[0,t]$. Therefore, the probability $u_k(t)$ satisfies
\begin{equation} \label{eq: uk}
\begin{aligned}
	u_{k + 1}(t) &= \int_0^t u_k(t-\tau)\psi(\tau)\mathrm{d}\tau\,,~~ k\ge 0\,.
\end{aligned}
\end{equation}
For any function $f: \mathbb{R}^N \longrightarrow \mathbb{R}$, let $\mathbb{E}[f]$ denote the expectation of $f(\bm x)$. Armed with this notation, we write
\begin{equation}
	\mathbb{E}[f](t) = \mathbb{E}[f({\bm x}(t))]\,.
\end{equation}
We take this expectation with respect to all sources of randomness except for the initial opinions. Let $\bm x[k]$ denote the opinion vector after $k$ updates, and let $\mathbb{E}_k[f]$ be the expected value of $f(\bm x[k])$. The event times are independent of opinion updates, so
\begin{equation} \label{eq: HK_approx}
	\mathbb{E}[f](t) = \sum_{k =0}^{\infty} \mathbb{E}_k[f]u_k(t)\,.
\end{equation}
The probability $u_k(t)$ is determined solely by the ITD $\psi$; it is independent of the update rules \eqref{eq: HK-random} and \eqref{eq: DW-random}.
The expectation $\mathbb{E}_k[f]$ is independent of both the ITD $\psi$ and the renewal process $R(t)$; it is determined solely by the update rules \eqref{eq: HK-random} and \eqref{eq: DW-random}.
Using the expression \eqref{eq: HK_approx}, we disassociate the expected opinion dynamics from the temporal stochasticity that arises from random-time interactions.
By introducing a cutoff for $k$, equation \eqref{eq: HK_approx} yields an approximate formula to compute the expected dynamics of our BCMs with random-time interactions.

We compute the probability $u_k(t)$ either directly using \eqref{eq: uk} or by employing the Laplace transforms of $u_k(t)$ to circumvent calculating the convolution. See \cite{chu2023nonmarkovian,hoffmann2012generalized} for how to derive the Laplace transforms of the probability $u_k(t)$.
The synchronous single-process BCM has a deterministic update rule \eqref{eq: HK-random}. 
Therefore, $\mathbb{E}_k[f] = f(x[k])$ and we obtain $\mathbb{E}_k[f]$ in \eqref{eq: HK_approx} with a single simulation of the discrete-time HK BCM \eqref{classical}. That is, we simulate ``one realization" of the discrete-time HK BCM \eqref{classical}.
For the asynchronous single-process BCM, it is often challenging to evaluate $\mathbb{E}_k[f]$ due to the randomness in selecting node pairs for potential opinion updates.
This randomness can yield different opinion trajectories for any ITD (even for the Dirac delta ITD). Therefore, we need to simulate multiple realizations of the discrete-time directed DW BCM \eqref{eq: DW_classical} to approximate the expectation $\mathbb{E}_k[f]$.

To quantify the amount of consensus in a simulation of a single-process BCM, we calculate the order parameter 
\begin{equation} \label{eq: Q}
	Q(\bm x) = \frac{1}{|E|}\sum_{e_{ij}\in E} \mathbbm{1}_{x_i = x_j} \,.
\end{equation}
When $Q = 1$, all agents have the same opinion and the system is in its most ordered phase. Conversely, when $Q = 1/N$ (where $N$ is the number of agents), each agent has a different opinion, so the system is in its least ordered phase. In practice, we often relax the condition $\mathbbm{1}_{x_i = x_j}$ by instead using $\mathbbm{1}_{|x_i - x_j| < \texttt{tol}}$ (where $\texttt{tol}$ is a tolerance parameter) to hasten the convergence of simulations. 

\begin{figure}[htp]
    \centering
    \includegraphics[width=0.23\textwidth]{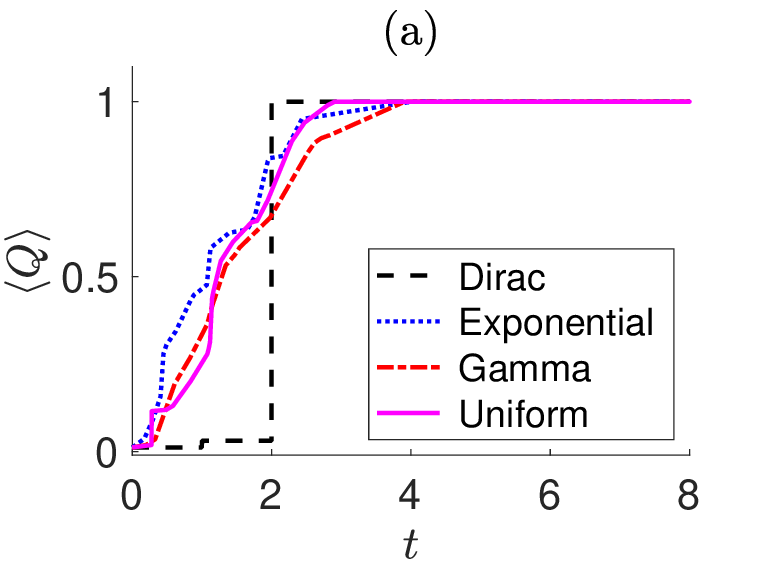}
    \includegraphics[width=0.23\textwidth]{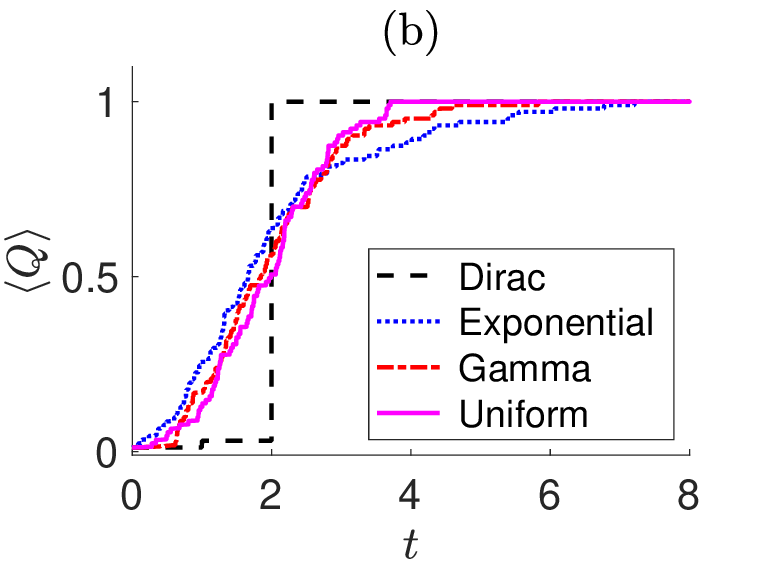} \vspace{8pt} \\
    \includegraphics[width=0.23\textwidth]{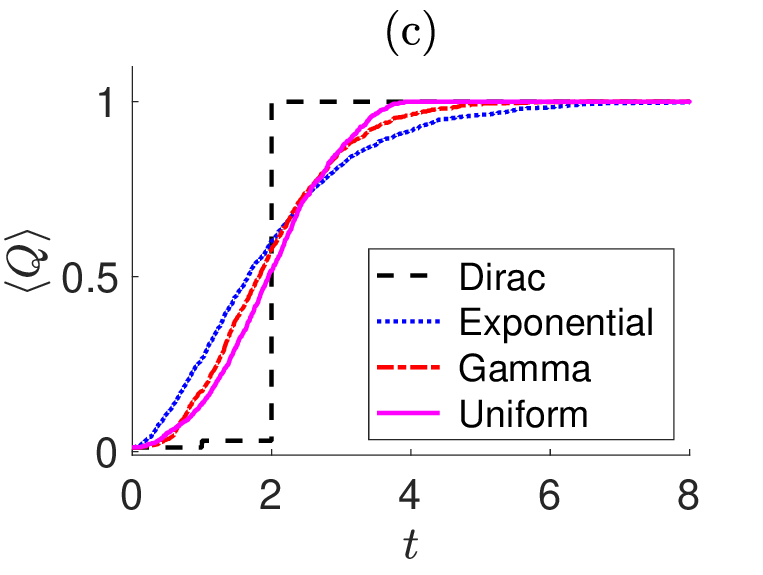}
    \includegraphics[width=0.23\textwidth]{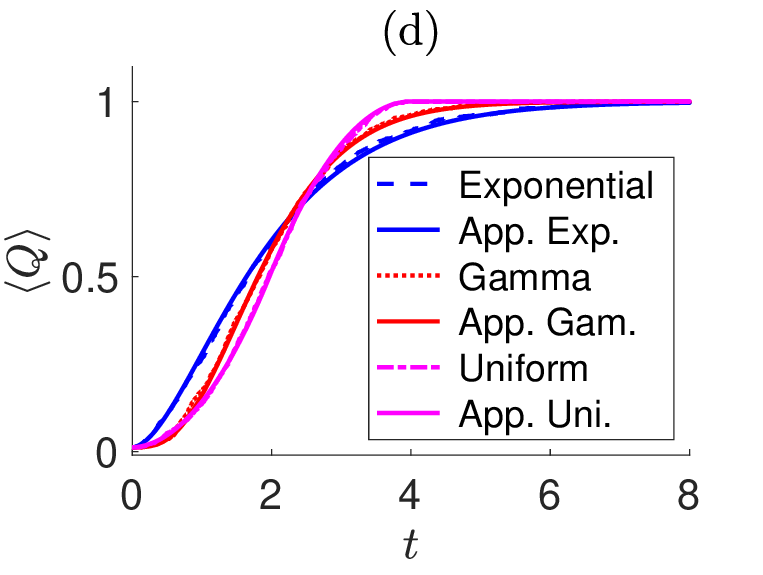}
    \caption{In (a)--(c), we show the sample means of the order parameter $Q$ [see \eqref{eq: Q}] of (a) $10$, (b) $100$, and (c) $1000$ simulations of the synchronous single-process BCM \eqref{eq: HK-random} on a 100-node complete graph. For each simulation, we draw the initial opinions from the uniform distribution on $[0,1]$. The confidence bound is $c = 0.5$, and the tolerance parameter is $\texttt{tol} = 10^{-2}$. In (d), we plot the sample means of $Q(\bm x)$ from (c) for different ITDs and their approximations using \eqref{eq: HK_approx}. In these approximations, we use $15$ as the upper bound of $k$. }
    \label{fig: SP-expectation}
\end{figure}

In Figure \ref{fig: SP-expectation}, we compute the mean of the order parameter $Q(\bm x)$ for the synchronous single-process BCM \eqref{eq: HK-random} for different ITDs and approximate the expected order parameters for the same models using \eqref{eq: HK_approx}. 
We distinguish between sample means $\langle \cdot \rangle$ and expectations $\mathbb{E}[\cdot]$ of quantities.
We consider renewal processes $R(t)$ with the ITDs
\begin{subequations} \label{eq: Ts}
\begin{align} 
    \psi_\text{Dirac}(t) &= \delta(t - \mu)\,,  \label{eq: Tdelta}\\
    \psi_\text{exponential}(t) &= \frac{1}{\mu}\exp(-t/\mu) \,, \label{eq: Texp} \\
    \psi_\text{gamma}(t) &= \frac{4t}{\mu^2}\exp(-2t/\mu)\,, \label{eq: Tgamma} \\
    \psi_\text{uniform}(t) &= \mathbbm{1}_{[0,2\mu]}(t)\,,  \label{eq: Tuniform}
\end{align}
\end{subequations}
where $\mathbbm{1}_S$ denotes the indicator function on the set $S$. All ITDs have the same mean value $\mu$.
As we increase the number of simulations, we observe that the time-dependent order parameters become smoother for the continuous ITDs (i.e., the exponential, Gamma, and uniform ITDs) and that the trajectories of the approximate expected order parameters closely match the trajectories of the mean order parameters for all ITDs.

\section{Multiple-process BCMs} \label{two}

The single-process BCMs in Section \ref{one} assume that a single renewal process governs the times of the interactions between agents.
In reality, however, people exchange opinions at various times, so one cannot expect the interactions between agents to be governed by a single renewal process. 
Therefore, we consider multiple independent renewal processes $R_{ij}$ (for all agents $i, j \in \{1, \ldots, N\}$.
The events in $R_{ij}$ trigger potential opinion updates of agent $i$ and determine when it interacts with each agent $j$.


\subsection{Interactions that are induced by multiple renewal processes}

Let $R_{ij}$ be a renewal process that generates a sequence $\mathcal{T}_{ij} = \{t_0,t_1,t_2,\ldots\}$ of event times with initial time $t_0 = 0$.
We suppose that all renewal processes $R_{ij}$ are independent of each other and have the same ITD $\psi$. The renewal process $R_{ij}$ determines the interaction times from agent $j$ to agent $i$.
At time $t\in\mathcal{T}_{ij}$, we update the opinion $x_i$ of node $i$ using the update rule\footnote{This is the same update rule as \eqref{eq: DW-random}, which we repeat for clarity.}
\begin{equation}\label{eq: multiple-process-BCM}
	x_i(t) = 
	\begin{cases}
	   	 \frac{1}{2} \left[x_i(t_-) + x_j(t_-)\right] &\text{if~} \left|x_i(t_-) - x_j(t_-)\right| < c   \\
	 	x_i(t_-) &\text{otherwise}\,.
	\end{cases}	
\end{equation}
The opinions of all other agents stay the same. If multiple events that involve the same agent $i$ occur simultaneously at time $t$, then we update its opinion $x_i$ to
\begin{equation} \label{eq: multiple-process-HK}
	 x_i(t) = \frac{\sum_{j\in \tilde{\mathcal{N}}_i(t_-)} x_j(t_-)}{|\tilde{\mathcal{N}}_i(t_-)|}\,, 
\end{equation}
where 
\begin{equation}
	\tilde{\mathcal{N}}_i(t) = \left\{ i \right\} \cup \left\{j \in {\mathcal{N}}_i(t): ~t\in \mathcal{T}_{ij} \right\}
\end{equation}
is a restricted neighbor set (which differs from the neighbor set \eqref{eq: neighbors}) that includes all neighboring nodes of $i$ that (1) interact with node $i$ at time $t$ and (2) have an opinion that differs from the opinion $x_i$ by less than the confidence bound $c$. 
We refer to \eqref{eq: multiple-process-BCM} as a \emph{multiple-process BCM}. 
When the ITD $\psi$ is continuous, the events of two renewal processes occur simultaneously with $0$ probability, so opinion updates in \eqref{eq: multiple-process-BCM} are asynchronous almost surely (i.e., with probability $1$).
When the ITD is $\psi(t) = \delta(t - \Dt)$ (i.e., the Dirac delta distribution), the events of different processes occur simultaneously at times $t = \Dt, \,2\Dt, \, \ldots$, and we obtain the synchronous single-process BCM in Section \ref{sec: HK}. We can extend the multiple-process BCM \eqref{eq: multiple-process-BCM} to a heterogeneous scenario in which each renewal process $R_{ij}$ has a different ITD $\psi_{ij}$. In such a model, opinion updates can occur as a hybrid of synchronous and asynchronous updates.

For the single-process BCMs \eqref{eq: HK-random} and \eqref{eq: DW-random}, the steady-state behaviors are statistically the same as in the classical BCMs \eqref{classical} and \eqref{eq: DW_classical}, respectively, as the random interevent times do not affect the order of agent interactions.
By contrast, in multiple-process BCMs, multiple renewal processes determine both the event times and the order of agent interactions.
Therefore, the steady-state behaviors are now statistically different from those in the classical BCMs.
We illustrate this situation using a 3-node network.

Consider a network with node set $V = \{1,2,3\}$ and edge set $E = \{e_{12}, e_{23}\}$. 
Suppose that the initial agent opinions are $x_1(0) = 0$, $x_2(0) = 0.5$, and $x_3(0) = 1$ and that the confidence bound is $c = 0.6$. 
By construction, the system achieves a steady state after a single opinion compromise. The steady state is $(x_1^*, x_2^*, x_3^*) = (0.25, 0.25, 1)$ if agents 1 and 2 compromise, and the steady state is $(x_1^*, x_2^*, x_3^*) = (0, 0.75, 0.75)$ if agents 2 and 3 compromise. 
Therefore, the system's steady state is determined entirely by whether agents 1 and 2 interact before agents 2 and 3 interact.
We consider two different choices of ITDs for the edges $e_{12}$ and $e_{23}$. In the first scenario, both edges follow the same ITD. Therefore, agent 2 interacts with agent 1 or agent 3 with equal probability. We thus obtain the steady state
\begin{equation}
	(x_1^*,x_2^*,x_3^*) = 
		\begin{cases}
    			(0.25,0.25,1) \quad \textrm{with probability} ~~0.5 \\
    			(0,0.75,0.75) \quad \textrm{with probability} ~~0.5 \,.
		\end{cases}
\end{equation}
In the second scenario, edge $e_{12}$ follows the uniform ITD on $[0.5,1.5]$ and edge $e_{23}$ follows the ITD 
\begin{equation}
	\psi_{23}(t) = \frac{3}{5}\delta(t - \frac13) + \frac{2}{5}\delta(t - 2)\,,
\end{equation}
which is a sum of two Dirac delta distributions. 
Therefore, when the interevent time of $e_{23}$ is $1/3$, agent 2 interacts with agent 3 before it interacts with agent 1 because the interevent time of $e_{12}$ is at least $1/2$, which is larger than $1/3$.
With probability $3/5$, agents 2 and 3 interact and thereby yield the steady state $x^* = (0,0.75,0.75)$.
Similarly, when the interevent time of $e_{23}$ is $2$, agent 2 interacts with agent 3 after it interacts with agent 1 because the interevent time of $e_{12}$ is at most $3/2$, which is smaller than $2$. 
With probability $2/5$, agents 1 and 2 interact and thereby yield the steady state $x^* = (0.25,0.25,0)$.
We thus obtain the steady state
\begin{equation}
	(x_1^*,x_2^*,x_3^*) = 
		\begin{cases}
			    (0.25,0.25,1) \quad \textrm{with probability} ~~{2/5} \\
			    (0,0.75,0.75) \quad \textrm{with probability} ~~{3/5}\,.
		\end{cases}
\end{equation}

In both scenarios, the ITDs of all edges have the same mean (which is equal to $1$). However, the two scenarios yield different steady-state statistics. 
This discrepancy arises because the order of edge events (and hence the sequence of pairwise interactions) determines the overall dynamics and is not statistically equivalent for different ITDs even when they have identical means.
Accordingly, in multiple-process BCMs (and unlike in single-process BCMs), the randomness in interaction times affects steady-steady behavior.
In Sections \ref{sec: markovian_multi} and \ref{sec: nonmarkovian_multi}, we examine how the types of ITDs and their parameters and network structures affect the dynamics of multiple-process BCMs.
However, for Markovian ITDs (such as the exponential and Dirac delta distributions), the expected opinions follow the same dynamics for ITDs with the same mean.


\subsection{Dynamics of Markovian multiple-process BCMs}
\label{sec: markovian_multi}

We now discuss the dynamics of some Markovian multiple-process BCMs. 

When the ITD is a Dirac delta distribution, both the single-process BCM \eqref{eq: HK-random} and the multiple-process BCM \eqref{eq: multiple-process-BCM} become discrete-time Markovian processes. In this situation, both models reduce to the classical HK BCM \cite{hegselmann2002opinion}. In the rest of this subsection, we consider continuous-time Markovian BCMs with an exponential ITD. To help highlight their dynamics, we also compare them to BCMs with a Dirac delta ITD.

When the ITD $\psi(t)$ is exponential, the renewal processes $R_{ij}$ are Poisson point processes. 
We write $\psi(t) = \lambda e^{-\lambda t}$, where $\lambda$ is the rate parameter of the process.
The sum (i.e., ``superposition") of $|E|$ Poisson point processes is a Poisson point process $P(t)$ with rate parameter $\Lambda = \lambda |E|$. 
In this case, the multiple-process BCM is the same as the asynchronous single-process BCM \eqref{eq: DW-random} with an exponential ITD with decay rate $\Lambda$. We show that the opinion model that is induced by the exponential ITD is Markovian, and we relate the dynamics of the expected opinions to a continuous-time HK BCM \cite{blondel2010continuous}.

Let $P(t)$ be the superposition of all $|E|$ Poisson point processes $R_{ij}$ (where $E$ is the set of edges of a network), and let $Z$ denote the total number of events in the time interval $[t,t + \tau)$ for $P(t)$. We have
\begin{equation}
    Z = \begin{cases}
	    0 & \text{with probability} \quad e^{-\Lambda\tau} \\
	    1 & \text{with probability} \quad \Lambda \tau \, e^{-\Lambda \tau} \\
		\ge 2 & \text{with probability} \quad \sum_{k = 2}^\infty \frac{(\Lambda \tau)^k}{k!} e^{-\Lambda \tau}  \,.
    \end{cases}
\end{equation}
When $Z = 0$, no opinion update occurs in the time interval $[t, t + \tau)$, so this situation does not contribute to opinion updates. 
When $Z = 1$, one event occurs in the time interval $[t, t + \tau)$. This event is generated by the process $R_{ij}$ with probability $1/|E|$. 
In this event, agent $i$ changes its opinion by
\begin{equation}
    \triangle_{i,j}(t) = \frac12\mathbbm{1}_{|x_i(t) - x_j(t)| < c}\left[x_j(t) - x_i(t)\right]\,.
\end{equation}

Let $y_i(t) = \mathbb{E}[x_i(t)]$ be the expectation with respect to the point-process superposition $P(t)$. The expected opinion of $y_i(t  + \tau)$ satisfies
\begin{equation} \label{eq: nonMarkovian}
	 y_i(t + \tau) =  y_i(t) + \!\!\sum_{{\{}j: e_{ij}\in E{\}}} \frac{\Lambda \tau}{|E|}\, e^{-\Lambda\tau} \mathbb{E}\left[\triangle_{i,j}(t)\right] + \mathcal{O}(\tau^2)\,,
\end{equation}
where the $\mathcal{O}(\tau^2)$ correction arises from the contribution for $Z \ge 2$. 
We use the relation $\Lambda = \lambda |E|$ and take the limit $\tau \to 0$ to obtain
\begin{equation} \label{eq: expected-exp}
		 \dot{y}_i(t) = \frac{\lambda}{2} \sum_{{\{j: e_{ij}\in E\}}} \!\!\mathbb{E}\left\{\mathbbm{1}_{|x_i(t) - x_j(t)| < c}\left[x_j(t) - x_i(t)\right]\right\}\,,
 \end{equation}
where $i \in \{1, \ldots, N\}$. 
The system \eqref{eq: expected-exp} is not closed. We make the bold approximation 
\begin{widetext}
\begin{equation}\label{eq: exp-approx}
	   \mathbb{E}\left\{\mathbbm{1}_{|x_i(t) - x_j(t)| < c}\left[x_i(t) - x_j(t)\right]\right\} \\
    \approx \mathbbm{1}_{|y_i(t) - y_j(t)| < c} \left[y_i(t) - y_j(t)\right]
\end{equation}
\end{widetext}
and insert \eqref{eq: exp-approx} into \eqref{eq: expected-exp} to obtain a closed set of equations. We thereby obtain
\begin{equation} \label{eq: HK-continuous}
	\dot{y}_i(t) = \frac{\lambda}{2}\sum_{{\{j: e_{ij}\in E\}}} \mathbbm{1}_{|y_i(t) - y_j(t)| < c} \left[y_j(t) - y_i(t)\right]\,,
\end{equation}
which is a continuous-time HK BCM \cite{blondel2010continuous} with $\lambda = 2$ and $c = 1$.
The approximation \eqref{eq: exp-approx} is the special form of the approximation $\mathbb{E}[g(r)] \approx g(\mathbb{E}[r])$ when $g(r) = \mathbbm{1}_{|r| < c} r$ and $r = x_i - x_j$. 
For a general random variable $r$, the expectation $\mathbb{E}[g(r)]$ does not equal $g(\mathbb{E}[r])$. These two quantities are equal in two special cases: (1) when $g$ is linear; and (2) when $r$ follows a Dirac delta distribution.  
In our numerical simulations, we observe discrepancies between \eqref{eq: expected-exp} and \eqref{eq: HK-continuous}. 

The expected dynamics \eqref{eq: expected-exp} is related to the asynchronous single-process BCM \eqref{eq: DW-random} when the ITD is the Dirac delta distribution $\psi(t) = \delta(t - \Dt)$ and $\Dt = 1/(\lambda|E|)$.
In this case, the agent opinions update at discrete times $t = \Dt,\,2\Dt,\,\ldots$\,. We consider a piecewise-linear interpolation of the opinions $x_i(t)$ on each time interval $[k\Dt,(k + 1)\Dt)$ with the opinions at the two interval endpoints. 
The expected opinions satisfy
\begin{equation}\label{eq: linear-piecewise}
	\dot{y}_i(t) = \frac{\lambda}{2} \!\sum_{{\{j: e_{ij}\in E\}}} \!\!\!\mathbb{E}\left\{ \mathbbm{1}_{|{r_{ji}(k\Dt)}| < c}{r_{ji}(k\Dt)}\right\}
\end{equation}
for $t \in [k\Dt, (k + 1)\Dt)$, where $r_{ji}(k\Dt) = x_j(k\Dt) - x_i(k\Dt)$. 
Equation \eqref{eq: linear-piecewise} gives the expected dynamics of the single-process BCM \eqref{eq: DW-random} with a Dirac delta ITD, and it is also a discrete-time version of the expected dynamics \eqref{eq: expected-exp} for single-process \eqref{eq: DW-random} and multiple-process \eqref{eq: multiple-process-BCM} BCMs with exponential ITDs.\footnote{See \cite{fennell2016limitations} for a discussion of continuous-time and discrete-time approximations of Markovian dynamics.}
We observe numerically that the expected dynamics \eqref{eq: expected-exp} and \eqref{eq: linear-piecewise} yield the same dynamics when we approximate the expectations using the empirical means of the time-dependent opinions.

\begin{figure}[htp]
    \centering
    \includegraphics[width=0.23\textwidth]{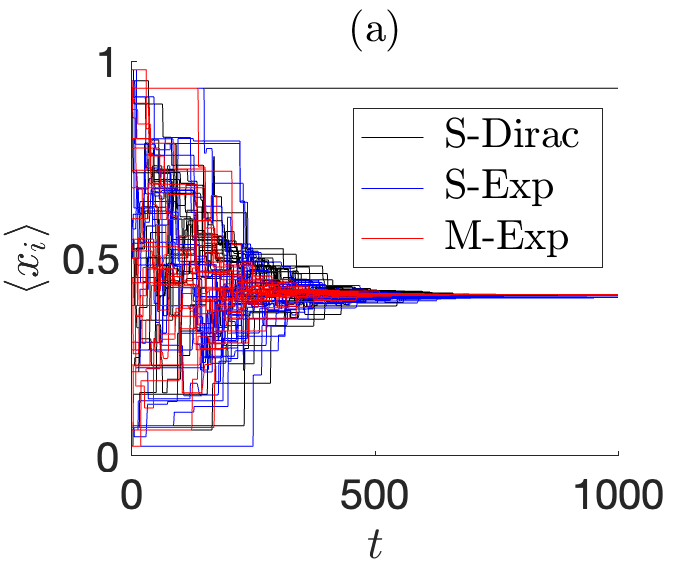}
    \includegraphics[width=0.23\textwidth]{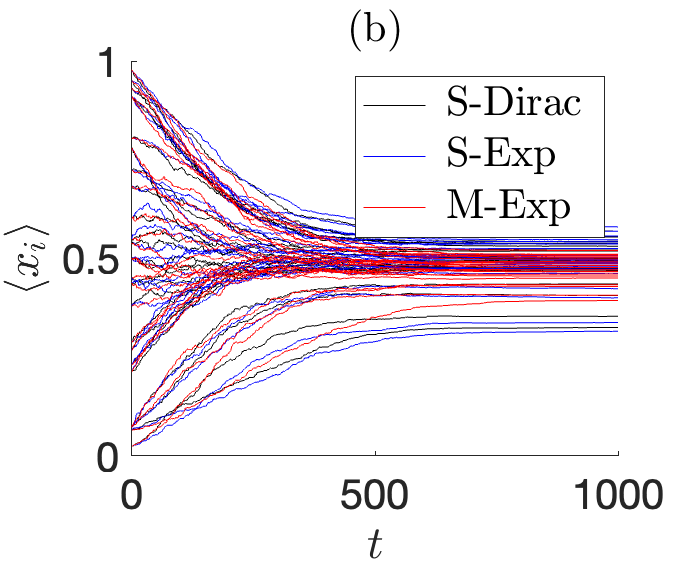} \\ \vspace{8pt}
    \includegraphics[width=0.23\textwidth]{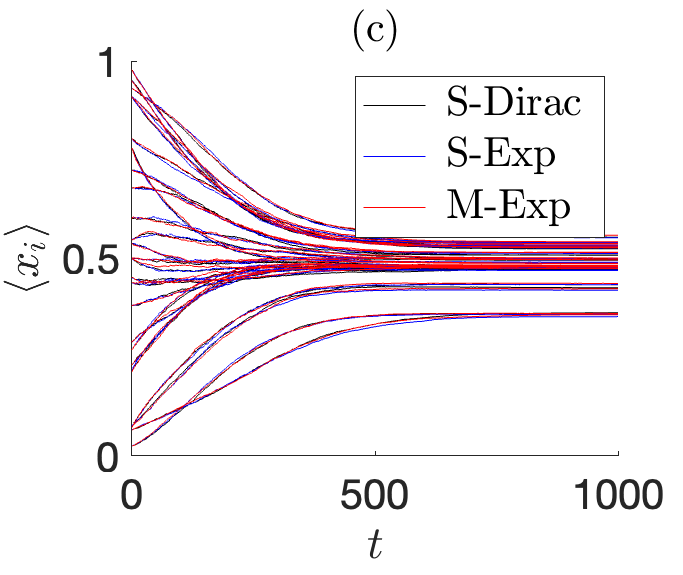}
    \includegraphics[width=0.23\textwidth]{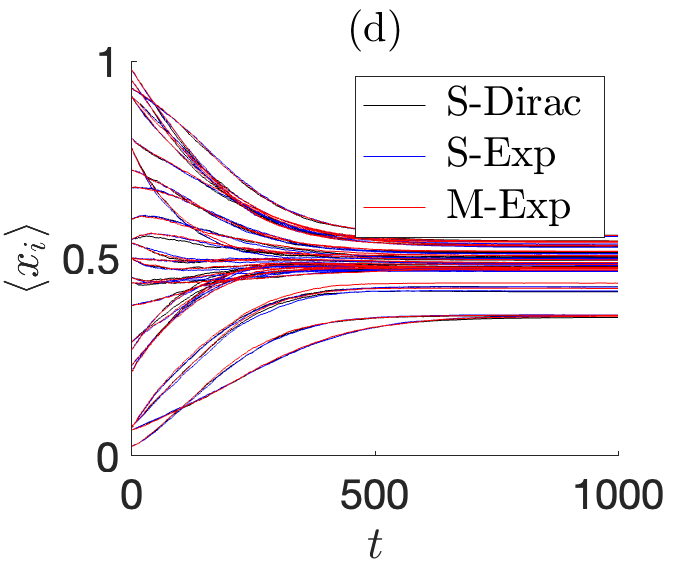}
    \caption{Sample means of the time-dependent opinions $x_i(t)$ in asynchronous single-process BCMs \eqref{eq: DW-random} with Dirac delta (S-Dirac) and exponential (S-Exp) ITDs and the multiple-process BCM \eqref{eq: multiple-process-BCM} with an exponential (M-Exp) ITD for (a) 1, (b) 100, (c) 1000, and (d) 2000 simulations.
    All simulations have the same initial opinions, which we draw uniformly at random from $[0,1]$. 
    We generate one directed 25-node $G(N,p)$ ER graph with connection probability $p = 0.5$, and we run all simulations on this ER graph.
    The confidence bound is $c = 0.4$. 
    }
    \label{fig: expected-exponential}
\end{figure}

In Figure \ref{fig: expected-exponential}, we show the mean time-dependent opinions of three Markovian models: the asynchronous single-process BCM \eqref{eq: DW-random} with the Dirac delta ITD $\psi(t) = \delta(t - 1/(\lambda|E|))$ (which we denote by ``S-Dirac''), the asynchronous single-process BCM \eqref{eq: DW-random} with the exponential ITD $\psi(t) = \lambda |E|\exp(-\lambda |E|t)$ (which we denote by ``S-Exp''), and the multiple-process BCM \eqref{eq: multiple-process-BCM} with the exponential ITD $\psi(t) = \lambda \exp(-\lambda t)$ (which we denote by ``M-Exp").
Each realization of these processes can have distinct opinion update times, so we use a piecewise-linear interpolation for opinions at discrete update times for each realization and then compute the mean of the interpolated opinion trajectories on the entire time domain. We then compute the mean opinion dynamics by averaging the interpolated dynamics across multiple simulations of the same model.

In our simulations, we observe that the expected dynamics are the same for our three Markovian BCMs, which are the models with single-process Dirac ITDs, single-process exponential ITDs, or multiple-process ITDs. 
Importantly, our observation that these three Markovian ITDs lead to the same expected opinion dynamics is independent of the network size $N$ and the confidence bound $c$. 
We use a small network (which has 25 nodes) because our simulations converge slowly to their expected values (due to the increase of the variance with network size).

\subsection{Gillespie algorithm for non-Markovian multiple-process BCMs}
\label{sec: nonmarkovian_multi}

It is computationally challenging to simulate a large number of processes in a multiple-process BCM \eqref{eq: multiple-process-BCM} with $|E|$ independent and concurrent renewal processes.
It is prohibitively complex to simulate these processes separately, organize their events chronologically, and execute opinion updates. To mitigate this computational burden, we use a Gillespie algorithm \cite{masuda2018gillespie}, which allows us to generate independent stochastic processes efficiently and statistically correctly.

\begin{algorithm}[H]
\caption{Gillespie algorithm to simulate $m$ independent renewal processes}\label{alg: GA}
\begin{algorithmic}[1]
\State Initialize $t_{\alpha} = 0$ for all $\alpha \in \{1,\ldots,m\}$.
\State Draw a uniform random variable $u$ from $[0,1]$ and determine the time increment $\Dt$ by solving
\begin{equation}
	\Phi\left(\Dt \mid\left\{t_{\alpha}\right\}\right) = \prod_{\alpha} \frac{\psi_{\alpha}(t_{\alpha} + \Dt)}{\Psi(t_{\alpha})} = u\,,
\end{equation}
where $\psi_{\alpha}$ is the ITD of the $\alpha$th renewal process and $\Psi_{\alpha}(t) = \int_t^{\infty} \psi_{\alpha}(\tau)\mathrm{d}\tau$ is the survival function.
\State Randomly select a process $\beta$ that generates an event with probability
\begin{equation}
	\Pi_{\beta} = \frac{\lambda_{\beta}\left(t_{\beta} + \Delta t\right)}{\sum_{\alpha = 1}^m \lambda_{\alpha}\left(t_{\alpha} + \Delta t\right)}\,,
\end{equation}
where 
\begin{equation}\label{eq: instantrate}
    \lambda_{\alpha}\left(t\right) = \frac{\psi_{\alpha}\left(t\right)}{\Psi_{\alpha}\left(t\right)}
\end{equation} 
is the instantaneous rate of the $\alpha$th process.
\State Set $t_{\beta} = 0$ and update $t_{\alpha}$ to $t_{\alpha} + \Delta t$ for $\alpha \neq {\beta}$.
\State Repeat steps 2--4 (or terminate the algorithm if a stopping criterion is satisfied).
\end{algorithmic}
\end{algorithm}

The traditional Gillespie algorithm \cite{gillespie1976general} is for independent Poisson processes, whose ITDs are exponential. Bogu{\~n}{\'a} et al.~\cite{boguna2014simulating} extended the Gillespie algorithm to simulate the events of multiple independent renewal processes. Their non-Markovian Gillespie algorithm draws a time increment $\Dt$ for the time to the next event from the superposition of $m$ renewal processes and determines the process that produces that event with a probability that depends on the interevent time of each renewal process. 
This non-Markovian Gillespie algorithm, which we state in Algorithm \ref{alg: GA}, generates a statistically correct sequence of event times.
One can terminate the algorithm after a specified number of events or when the time reaches a specified value.
When all renewal processes are Poisson processes, the instantaneous rate $\lambda_{\alpha}(t)$ in \eqref{eq: instantrate} reduces to the constant $\lambda_{\alpha}$, which is the rate of the $\alpha$th Poisson point process. 
That is, in this situation, this non-Markovian Gillespie algorithm reduces to the traditional Gillespie algorithm.

\begin{figure*}[hpt]
    \includegraphics[width=0.3\textwidth]{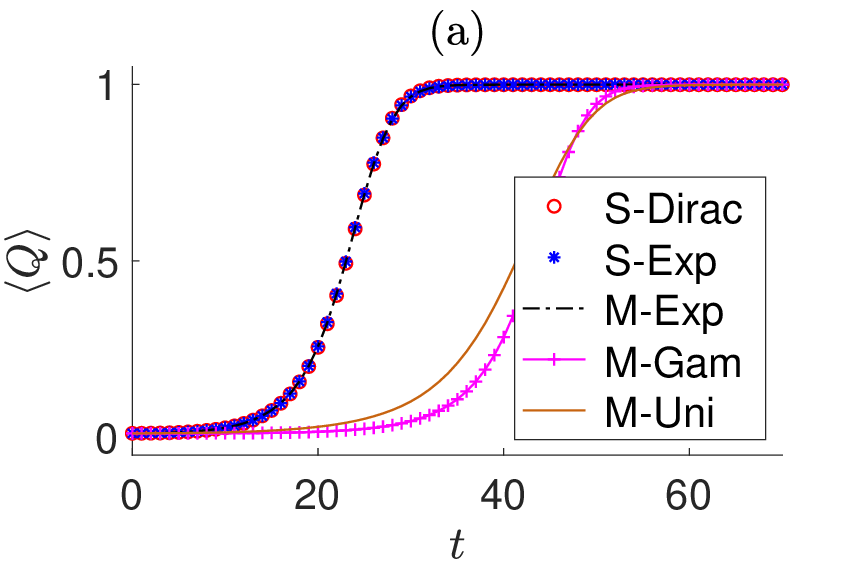}
    \includegraphics[width=0.3\textwidth]{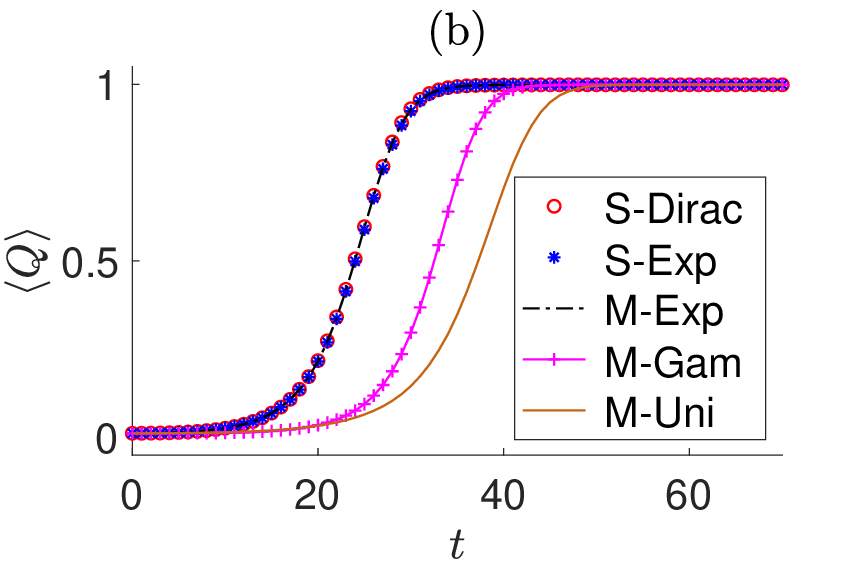}
    \includegraphics[width=0.3\textwidth]{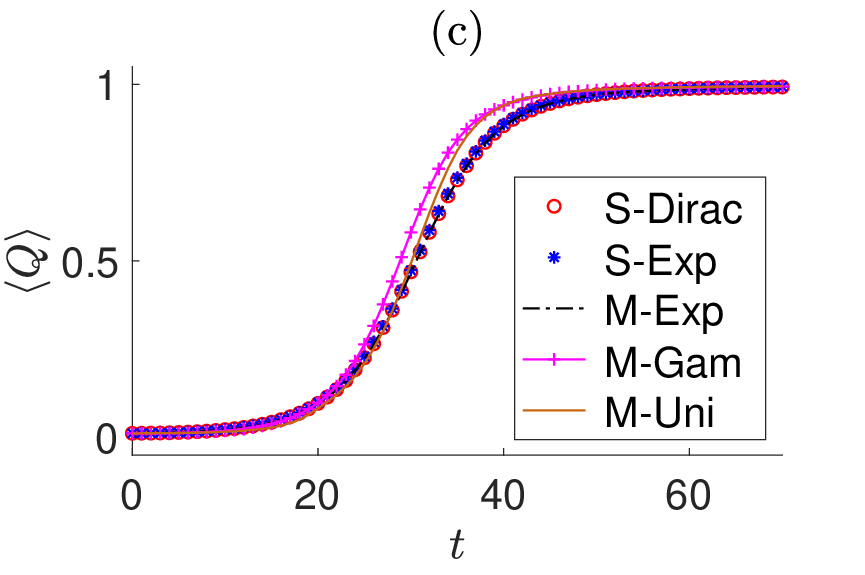} \vspace{5pt} \\
    \includegraphics[width=0.3\textwidth]{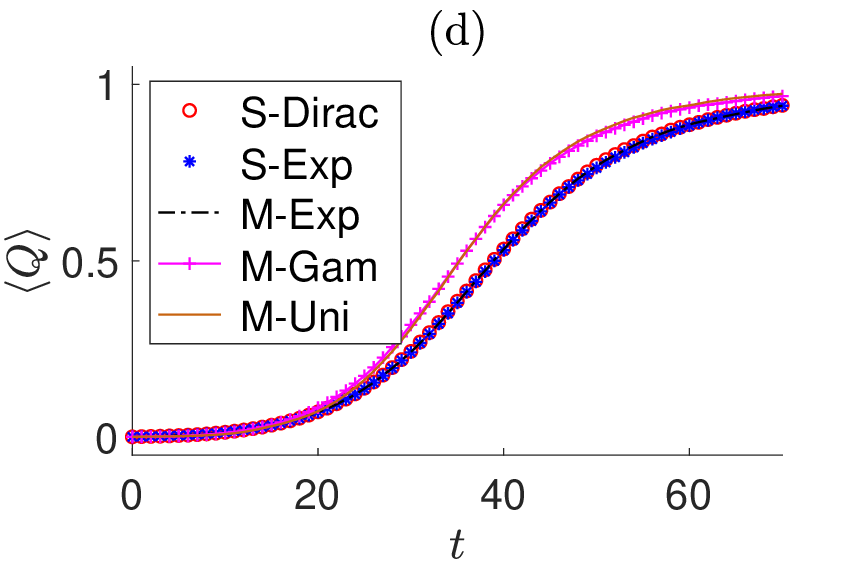}
    \includegraphics[width=0.3\textwidth]{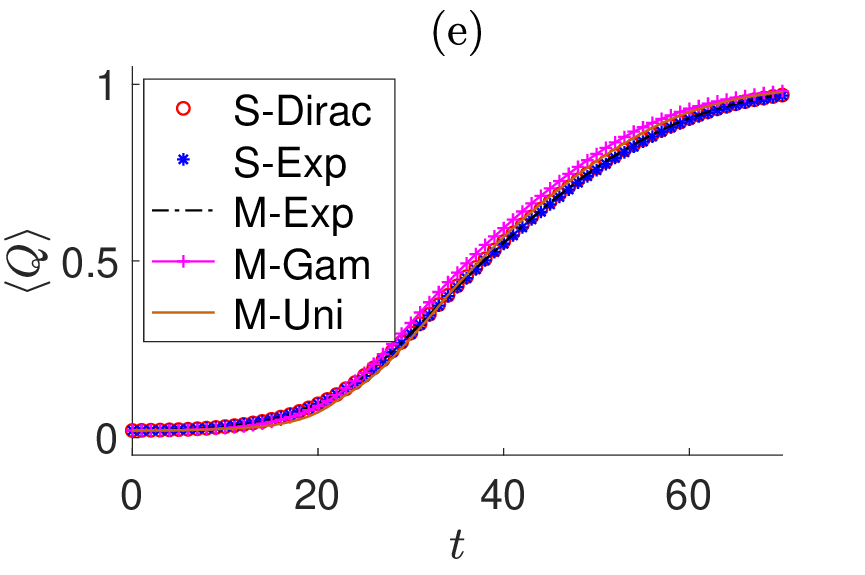}
    \includegraphics[width=0.3\textwidth]{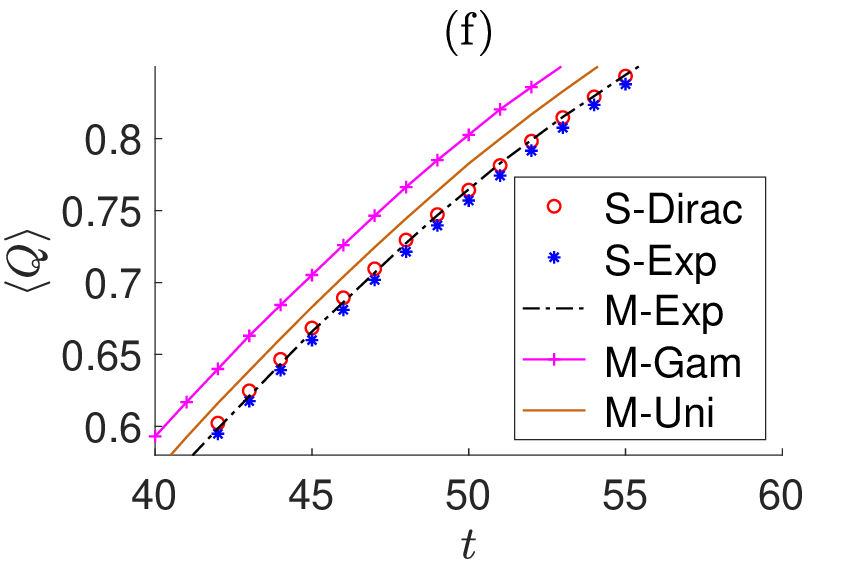}
    \vspace{5pt} \\
    \includegraphics[width=0.3\textwidth]{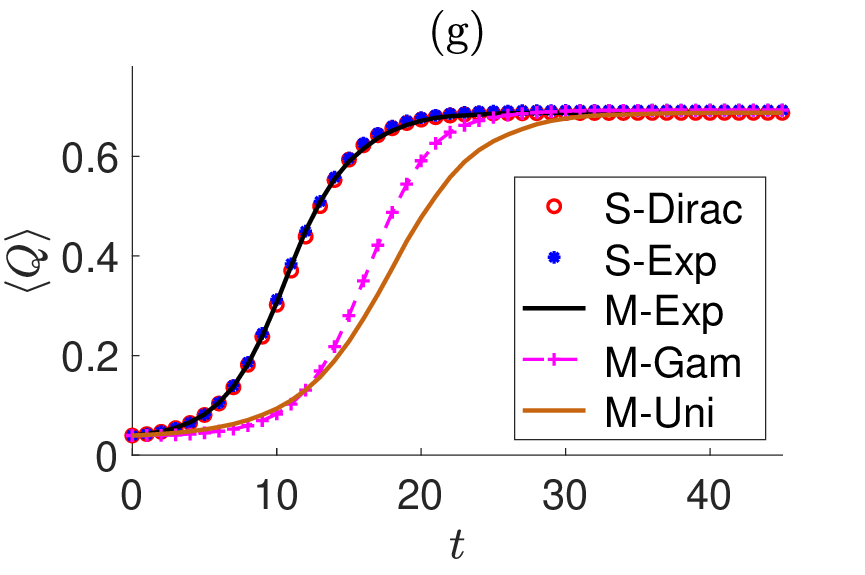}
    \includegraphics[width=0.3\textwidth]{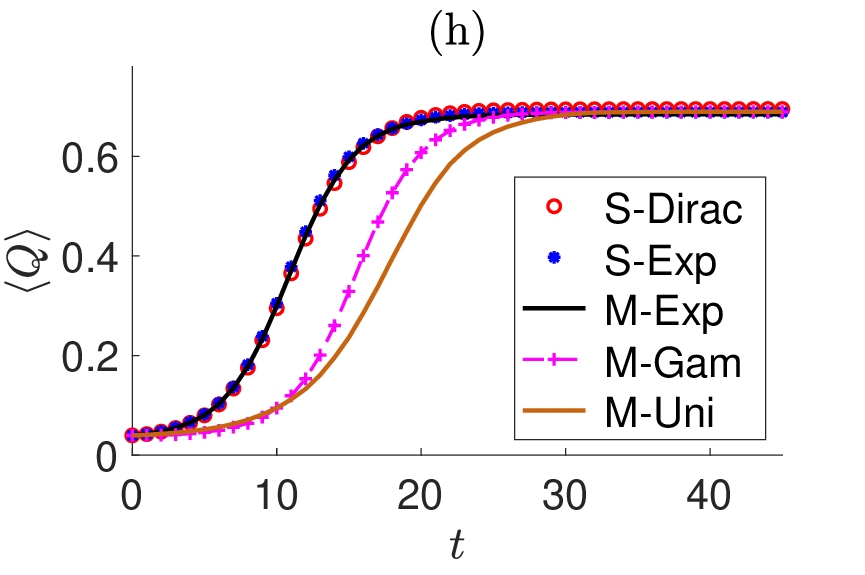}
    \includegraphics[width=0.3\textwidth]{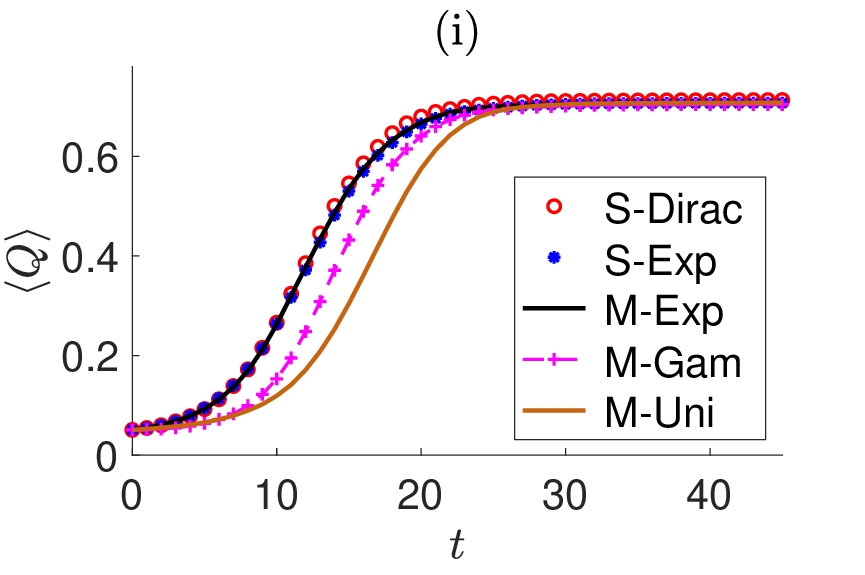}
    \caption{Sample means of the order parameter $Q$ [see \eqref{eq: Q}] versus time for single-process BCMs \eqref{eq: DW-random} with Dirac delta (S-Dirac) and exponential (S-Exp) ITDs and for multiple-process BCMs \eqref{eq: multiple-process-BCM} with exponential (M-Exp), gamma (M-Gam), and uniform (M-Uni) ITDs on (a,g) a complete graph, (b,h) directed $G(N,p)$ ER graphs with $p = 0.4$, (c,i) directed $G(N,p)$ ER graphs with $p = 0.1$, (d) symmetric and directed Chung--Lu graphs, and (e) directed SBM graphs with two communities. In (f), we show a magnification of (e).
    The ITD mean is $\mu = 0.01$. The confidence bound is $c = 0.5$ in panels (a)--(f) and is $c = 0.3$ in panels (g)--(i). 
    We compute the mean of $Q$ using 3000 BCM simulations. We draw the initial opinions uniformly at random from $[0,1]$ for each simulation, and we generate a new random graph for each simulation.
    }
    \label{fig: orderQ}
\end{figure*}

We use the non-Markovian Gillespie algorithm in Algorithm \ref{alg: GA} to simulate the multiple-process BCM \eqref{eq: multiple-process-BCM} on four distinct types of 50-node graphs: (1) a complete graph; (2) directed analogues of $G(N,p)$ ER random graphs in which each edge exists with independent and homogeneous probability $p$; (3) symmetric and directed Chung--Lu graphs (in which we start with undirected graphs and treat each undirected edge as two directed edges)~\cite{chung2002}, which are similar to configuration-model networks and are parametrized by sequences of expected degrees \cite{fosdick2018}, which we choose to be $\{10/\ln(k)\}_{k = 2,\ldots,N + 1}$; and (4) directed stochastic-block-model (SBM) graphs with two communities, intra-community probability $p_\text{AA} = p_\text{BB} = 0.2$, and inter-community probability $p_\text{AB} = p_\text{BA} = 0.02$. 
We explore how randomness, which arises through the ITDs and specific structures in the random-graph models, influences the order parameter $Q$ [see \eqref{eq: Q}] and convergence time in both single-process and multiple-process BCMs. 
For the single-process BCMs, we use Dirac delta and exponential ITDs with mean $\mu|E|$. Because the mean is the same, all simulations have the same expected number of events.
For the multiple-process BCMs \eqref{eq: multiple-process-BCM}, we consider exponential, gamma, and uniform ITDs with mean $\mu$.

In Figure \ref{fig: orderQ}, we plot the mean of the order parameter $Q$ [see \eqref{eq: Q}] from 3000 simulations of each scenario. We generate a new random graph for each simulation.
The three Markovian models --- the single-process BCM with the Dirac delta ITD, the single-process BCM with the exponential ITD, and the multiple-process BCM with exponential ITD --- yield almost identical mean time-dependent order parameters $Q$, which agrees with the results in Figure \ref{fig: expected-exponential}. We use piecewise-linear interpolation to construct $\mathbb{E}[Q^\text{Dirac}({\bm x}(t))]$ in the model with the Dirac delta ITD. Because of this construction, $\mathbb{E}[Q^\text{Dirac}({\bm x}(t))]$ is a piecewise-linear approximation of the value $\mathbb{E}[Q^\text{exp}({\bm x}(t))]$ that we obtain from the model with the exponential ITD. Equations \eqref{eq: expected-exp} and \eqref{eq: linear-piecewise} illustrate a similar relationship between the expected opinions. 
As $\triangle t \rightarrow 0$, we anticipate that the expected dynamics of both models converge to the same dynamics. In our simulations, we use $\triangle t = 0.01$. 
Multiple-process BCMs with gamma and uniform ITDs yield different dynamics for the mean of time-dependent order parameter $Q$, and the convergence rates depend both on the ITDs and on network structure.

Multiple-process BCMs with different non-Markovian ITDs can have distinct steady states, and their order parameters almost never converge to the same value (although they tend to yield similar values).
Moreover, when the confidence bound is $c = 0.5$, our simulations (of models with either Markovian or non-Markovian ITDs) do not always converge to a consensus, so the order parameter $Q$ does not converge exactly to 1. Instead, it converges to values less than 1.
When the confidence bound is $c = 0.3$, we observe polarization and segmentation in the steady states, with the order parameter again converging to a value less than 1.

\begin{figure*}[ht]
    \centering
    \includegraphics[height=0.19\textwidth]{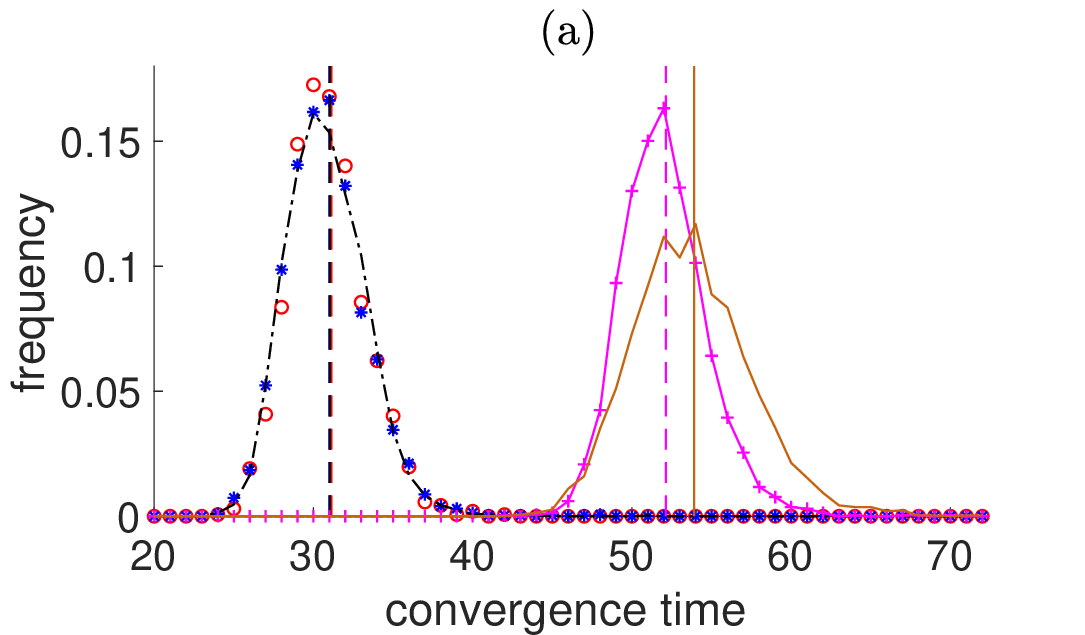}
    \includegraphics[height=0.19\textwidth]{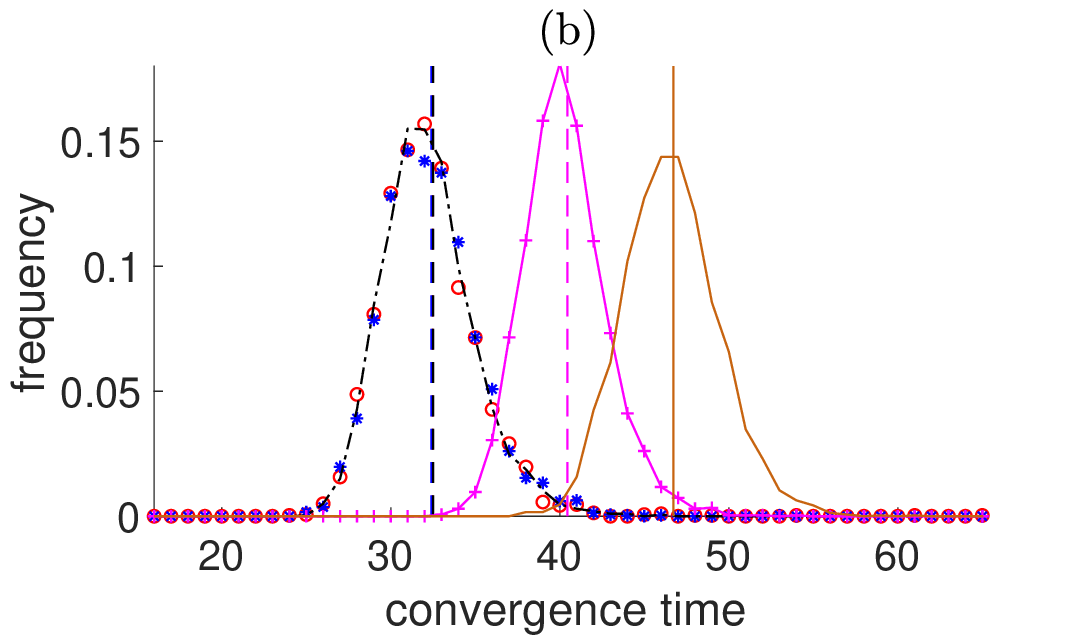}
    \includegraphics[height=0.19\textwidth]{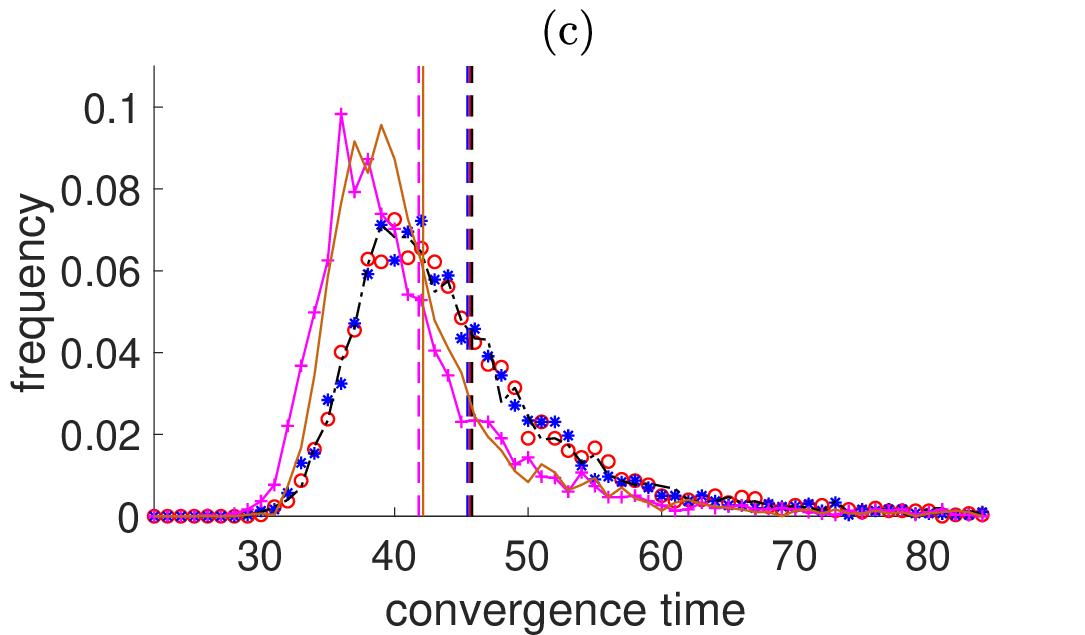} \vspace{10pt}\\
    \includegraphics[height=0.19\textwidth]{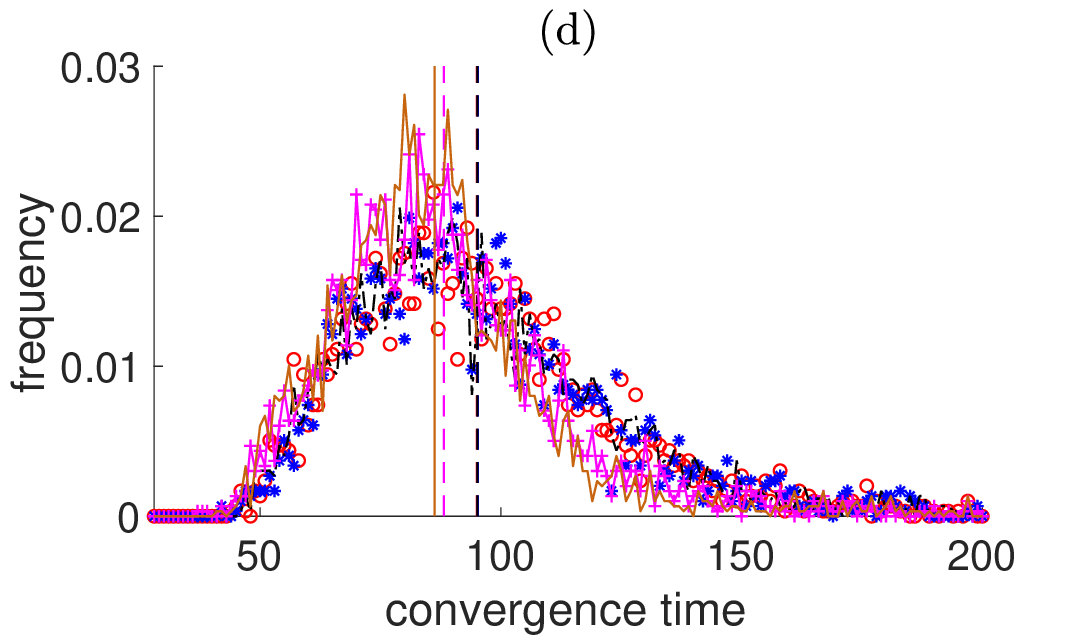}
    \includegraphics[height=0.19\textwidth]{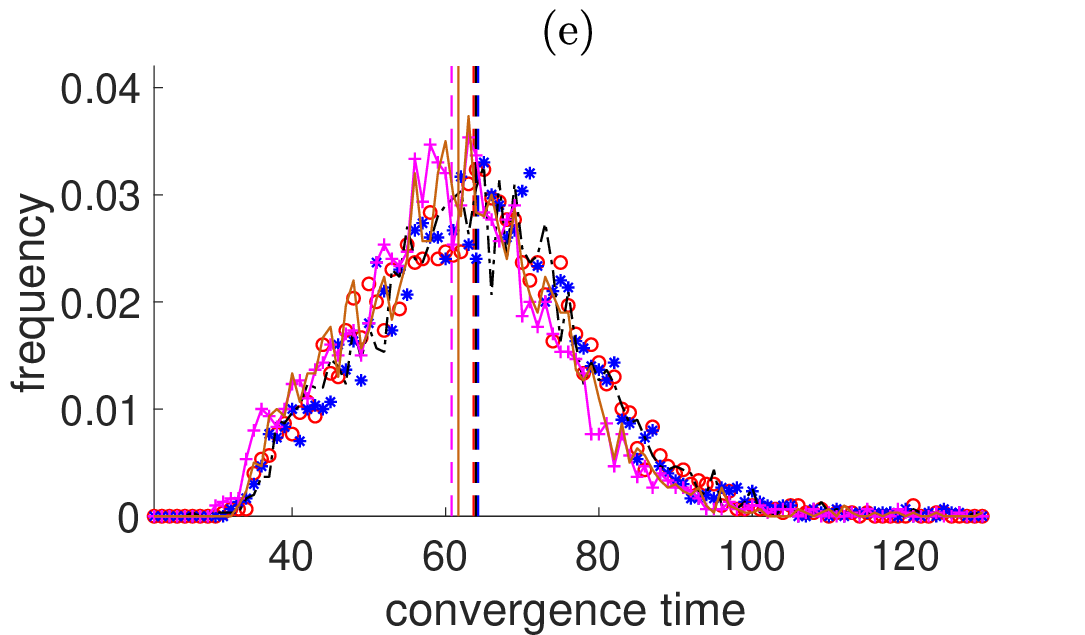}
~~~~\includegraphics[height=0.19\textwidth,trim={12.6cm 1cm 0 1cm},clip]{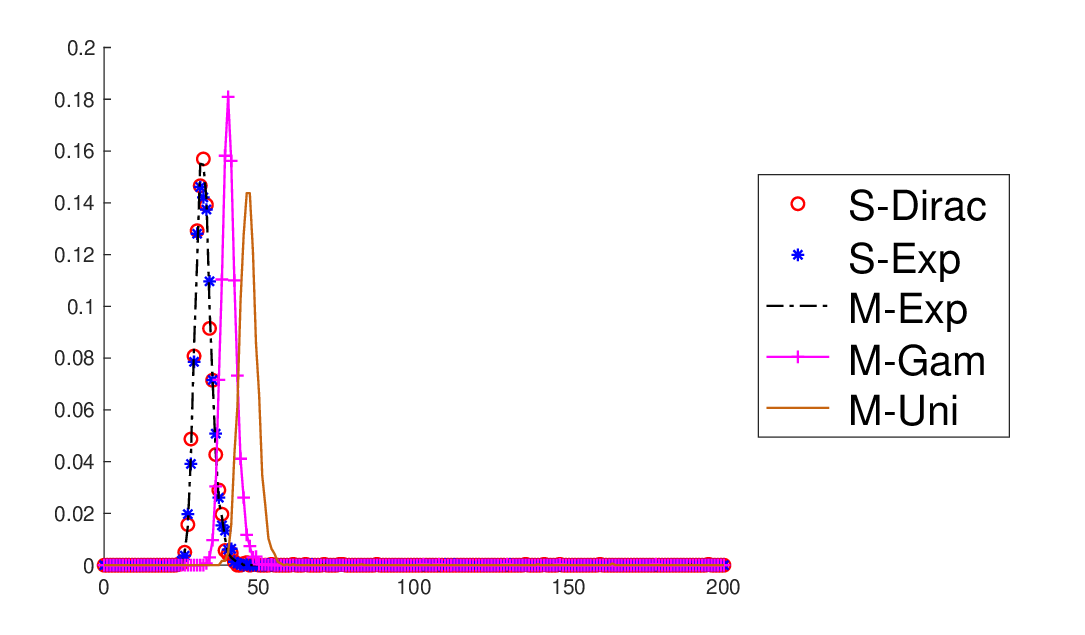}
    \caption{Normalized histograms of the convergence times of several of the simulations in Figure \ref{fig: orderQ}. The vertical lines indicate the mean convergence times of the BCMs on (a) a complete graph, (b) directed $G(N,p)$ ER graphs with 
    $p = 0.4$, (c) directed $G(N,p)$ ER graphs with $p = 0.1$, (d) symmetric and directed Chung--Lu graphs, and (e) directed SBM graphs with two communities. In panels (d) and (e), which show results for graphs with heterogeneous degree distributions, the variances of the convergence times are larger than those in panels (a)--(c). }
    \label{fig: convergence-time}
\end{figure*}

In Figure \ref{fig: convergence-time}, we show normalized histograms of the convergence times of the simulations in Figure \ref{fig: orderQ}. 
We treat a simulation as having converged if the opinion difference between each pair of adjacent nodes is either (1) at least the confidence bound $c$ or (2) smaller than $10^{-3}$. 
Based on our numerical observations, we see that both the ITD and network structure influence convergence time.
Additionally, as we increase the heterogeneity of a degree distribution, the convergence-time variance increases dramatically.


\section{Conclusions and discussion} \label{three}

Social systems include various elements of randomness, and it is important to account for such randomness in models of opinion dynamics.
In this paper, we extended classical bounded-confidence models (BCMs) of opinion dynamics by incorporating randomness into agent interaction times.
In our BCMs, opinion updates occur randomly in time as events of renewal processes. The interevent times are random and follow non-Markovian interevent-time distributions (ITDs). The classical Hegselmann--Krause (HK) and Deffuant--Weisbuch (DW) BCMs arise from specific choices of the renewal processes and are thus special cases of our models.
We investigated how ITDs affect the transient dynamics of our BCMs, and we derived approximate master equations to describe the time-dependent expectations of opinions. 
We numerically simulated our BCMs on various types of networks to explore how different network structures impact their dynamics.

In the single-process BCMs \eqref{eq: HK-random} and \eqref{eq: DW-random}, for which a single renewal process governs the interaction times between agents, the ITDs only influence the transient opinion dynamics. One obtains the same steady-state outcome for all ITDs. For a Dirac delta ITD, we highlighted that the model \eqref{eq: HK-random} reduces to the classical HK BCM \cite{hegselmann2002opinion} and that the model \eqref{eq: DW-random} reduces to a directed variant of the classical DW BCM \cite{deffuant2000mixing}.
Additionally, we derived a relationship \eqref{eq: HK_approx} between single-process BCMs (\eqref{eq: HK-random} and \eqref{eq: DW-random}) and the deterministic-time BCMs (\eqref{classical} and \eqref{eq: DW_classical}, respectively) in terms of their expected dynamics. 
This relationship \eqref{eq: HK_approx} yields an approximation method to efficiently compute the expected dynamics of the single-process BCMs.
Using numerical simulations, we demonstrated that our approximation is accurate for exponential, Gamma, and uniform ITDs.

We also developed multiple-process BCMs \eqref{eq: multiple-process-HK}, which use multiple independent renewal processes to determine the interaction times between agents.
Multiple-process BCMs with Dirac delta and exponential ITDs yield Markovian models that are equivalent to single-process BCMs with appropriately chosen ITDs and parameters. 
We derived an approximate governing equation for the expected opinions in these two Markovian models, and we showed that one can interpret the expected dynamics of the BCMs with Dirac delta ITDs as a discrete-time analogue of the expected dynamics of the BCMs with exponential ITDs.
For specific parameter values, these two models reduce to the continuous-time BCM in~\cite{blondel2010continuous}.
To numerically simulate our multiple-process BCMs efficiently and statistically accurately, we employed a non-Markovian Gillespie algorithm~\cite{boguna2014simulating}. In our numerical computations, we observed that both ITDs and network structure significantly influence the transient properties --- including both the order parameter \eqref{eq: Q} and the convergence times --- of the non-Markovian BCMs.
We also observed that network structures have a larger influence than the choice of ITDs on the convergence-time variance and that node heterogeneities further amplify this variance.

In the present paper, we examined BCMs on unweighted graphs and assumed that the ITDs are homogeneous across all edges. It is worthwhile to incorporate both heterogeneous edge weights and heterogeneous ITDs.
In such extensions, one can incorporate heterogeneity in a network's edges (as opposed to the node heterogeneities in Ref.~\cite{li2023}) and use weighted averages in synchronous opinion updates \eqref{eq: HK-random} or encode heterogeneous ITDs with parameters that are linked to edge weights. 
In most of our numerical simulations, we used a value of the confidence bound that typically leads to consensus.
To thoroughly study transitions between consensus steady states and other outcomes (especially polarization and fragmentation), it is important to also systematically examine many values of the confidence bound. 
It will be particularly interesting to explore the impact of different ITDs on the transitions between different steady states (such as between consensus and polarization) and on their convergence times. 
Another interesting research avenue is to incorporate temporal stochasticity into density-based BCMs~\cite{ben2003bifurcations,chu2023density}, which describe the collective behavior of a large population of agents and take the form of integro-differential equations. Naturally, it is also worth exploring the behavior of BCMs with random-time interactions on real social networks and with ITDs that one estimates from empirical data.


\begin{acknowledgements}

MAP was funded in part by the National Science Foundation (grant number 1922952) through their program on Algorithms for Threat Detection. WC was funded in part by the National Science Foundation through DMS-2514053.

\end{acknowledgements}



\bibliography{references}


\end{document}